\documentclass[preprint,aps,showpacs,nofootinbib,preprintnumbers,amsmath,amssymb]{revtex4-1}
\usepackage{epsfig}
\usepackage{subfigure}
\usepackage{dcolumn}
\usepackage{bm}
\usepackage[usenames ,dvipsnames]{xcolor}
\usepackage{slashed}
\usepackage{graphicx,color}
\usepackage{amsmath}
\usepackage{nonfloat}
\begin{document}

\title{Cosmological evolutions in Tsujikawa model of $f(R)$ Gravity}

\author{Jian-Yong Cen$^{1}$, Shang-Yu Chien$^{2}$\footnote{s104022511@m104.nthu.edu.tw}, Chao-Qiang Geng$^{2,3,1}$ and Chung-Chi Lee$^{4}$}
\affiliation{
	$^{1}$School of Physics and Information Engineering, Shanxi Normal University, Linfen 041004\\
	$^{2}$Department of Physics, National Tsing Hua University, Hsinchu 300\\
	$^{3}$Physics Division, National Center for Theoretical Sciences, Hsinchu 300\\
	$^{4}$DAMTP, Centre for Mathematical Sciences, University of Cambridge, Wilberforce Road, Cambridge CB3 0WA
}

\date{\today}

\begin{abstract}
We concentrate on the cosmological properties in the Tsujikawa model (TM) of viable $f(R)$ gravity with the dynamical background evolution
and  linear perturbation theory by using the CosmoMC package. We study  the constraints of the cosmological variables along with 
the model parameter from the current observational data. In particular, we show that 
the matter density fluctuation is suppressed and the constraint of the neutrino mass sum is relaxed  in the TM, which
are similar to other viable $f(R)$ models.
In addition, we discuss the parameters of  the deceleration and equation of state for dark energy in the TM
and compare them with those in the $\Lambda$CDM model.
\end{abstract}

\maketitle

\section{Introduction}\label{sec:1}

To describe the accelerating expansion of the universe, the  $\Lambda$CDM  model~\cite{Amendola:2015ksp}  is the simplest candidate.
However, this simplest version has the so called  ``cosmological constant problem,'' 
which is related to the ``fine-tuning''~\cite{Weinberg:1988cp, WBook} and ``coincidence''~\cite{Peebles:2002gy,Ostriker:1995rn,ArkaniHamed:2000tc}
 problems. People have been motivated by these issues  to explore new theories beyond $\Lambda$CDM, such as those with the dynamical dark energy~\cite{Copeland:2006wr,Li:2011sd}. A typical way is to modify the standard general relativity (GR) by promoting the Ricci scalar of $R$ 
in the Einstein-Hilbert action to an arbitrary function,   i.e., $f(R)$~\cite{DeFelice:2010aj}.
In addition,  many viable  $f(R)$ gravity models have been developed in the literature~\cite{DeFelice:2010aj} 
to satisfy the theoretical and observational constraints.
The most  popular  ones are  Hu-Sawicki~\cite{Hu:2007nk},  Starobinsky~\cite{Starobinsky:2007hu}, 
Tsujikawa~\cite{Tsujikawa:2007xu}, and  exponential~\cite{Exponential-type-f(R)-gravity,Cognola:2007zu, Linder:2009jz,Bamba:2010ws} 
$f(R)$ gravity models, in which the first three have been extensively examined in the literature, such as the recent one in Ref.~\cite{CQG:2019}, 
whereas the last one, $i.e.$ the Tsujikawa model (TM), has not been systematically explored yet, which is our concentration in this study. 

The TM  of the viable $f(R)$ models, first proposed by Tsujikawa in 2007~\cite{Tsujikawa:2007xu}, 
is written as
\begin{equation}
\label{Tsujikawa model}
f(R)=R-\lambda R_{ch}\tanh\left(\frac{R}{R_{ch}}\right)\,,
\end{equation}
where $\lambda$ is the dimensionless model-parameter and $R_{ch}$ corresponds to the constant characteristic curvature in the model.
The TM has a simpler form than most of other  f(R) models as it contains only one model-parameter beyond  $\Lambda$CDM,
which can be regarded as a similar type of the exponential $f(R)$ model~\cite{Ali:2010zx}, 
but has a different form in the functional structure, which could result in some different cosmological behaviors in the numerical results.

It is known that the viable $f(R)$ gravity can well describe the power spectrum of the matter density fluctuation~\cite{Yang:2010xq,Hu:2013twa, Raveri:2014cka, deMartino:2015zsa, Geng:2014yoa, Geng:2015vsa} and the formation of the large scale structure (LSS)~\cite{Li:2011vk, Puchwein:2013lza, Lombriser:2013wta, Llinares:2013jza} in the universe. To investigate the dynamical dark energy behaviors, we reply on  the existing  open-source programs.
However, most of them are written with either the parametrization in term of the equation of state or the background evolution being the same as 
the $\Lambda$CDM model~\cite{Geng:2014yoa, Geng:2015vsa}.
Recently, the allowed parameter spaces of the cosmological observables in the viable $f(R)$ gravity models
with the dynamical background evolution  have been explored in Ref.~\cite{CQG:2019}.
In this work, we will use the same method to examine the TM.
In particular, we will show  the allowed windows for the active neutrino masses, dark energy density  and Hubble parameter
as well as other cosmological parameters, such as the deceleration and equation of state for dark energy.
In addition, 
since the first detection (GW150914) of gravitational waves  by the LIGO Collaboration~\cite{Abbott:2016blz}, 
the gravitational radiation 
has been believed to be a new tool to test GR and search for  new physics. Beyond GR, the gravitational waveforms 
of the  modified gravity theories have been discussed in the literature~\cite{Liu:2018sia,Jana:2018djs,Kase:2018aps}.
It is possible that the TM and  other f(R) models may be stringently constrained by the future gravitational wave  detectors.

In this paper, we take  the open source program of  the Modification of Growth (MG) with 
Code for Anisotropies in the Microwave Background (CAMB)~\cite{Lewis:1999bs, Hojjati:2011ix}, 
which is designed to examine the dynamical dark energy model. In order to put the TM into the program of MGCAMB, 
 we modify the growth equations of the scalar perturbations and  density fluctuations in the Newtonian gauge.
We also include the dynamical background evolution of  dark energy~\cite{Geng:2014yoa, Geng:2015vsa} instead of the $\Lambda$CDM one, 
and use the MG Cosmological MonteCarlo (MGCosmoMC)~\cite{Lewis:2002ah,Zhao:2008bn} together with the latest data from
the cosmological observationas.

The paper is organized as follows. In Sec.~\ref{sec:2}, we present the TM of  viable  $f(R)$ gravity. 
In Sec.~\ref{sec:3}, we show the cosmological evolutions  in the TM. In particular, we include the perturbation 
equations of the dynamical background evolution. In Sec.~\ref{sec:4}, we show the constraints from the cosmological observational data. 
Finally, our conclusions are given in the Sec.~\ref{sec:5}.


\section{Tsujikawa model of viable $f(R)$ gravity}\label{sec:2}

The modified Einstein Hilbert action of the $f(R)$ gravity models is given by
\begin{equation}
\label{eq:action}
S=\int{ d^{4}x \frac{\sqrt{-g}}{2\kappa^2} f(R)}+S_{M}\,,
\end{equation}
where $\kappa^{2}=8\pi G$ with $G$  the Newton's constant, $g$ stands for the determinant of the metric tensor $g_{\mu\nu}$, 
$f(R)$ is an arbitrary function of the Ricci scalar $R$, 
and $S_{M}$ corresponds to the action of the relativistic and non-relativistic matter.
After the variation of $g_{\mu\nu}$ in the action, we obtain the modified field equation:
\begin{equation}
\label{eq:field}
f_R R_{\mu\nu}-\frac{f}{2}g_{\mu\nu}-\left(\nabla_{\mu}\nabla_{\nu}-g_{\mu\nu}\Box\right) f_R = \kappa^{2}T^{\left(M\right)}_{\mu\nu}\,,
\end{equation}
where $f_R \equiv df(R)/dR$, $\nabla_{\mu}$ is the covariant derivative,
 $\Box \equiv g^{\mu \nu}\nabla_{\mu} \nabla_{\nu}$ represents  the d'Alembert operator, 
and $T^{\left(M\right)}_{\mu\nu}$ denotes the energy momentum tensor.
To describe our universe, we use the Friedmann-Lema\"itre-Robertson-Walker (FLRW) metric, given by
\begin{equation}
\label{nonmetric}
ds^{2} = g_{\mu\nu}dx^{\mu}dx^{\nu} = -dt^{2}+a^{2} \left( t \right) d\vec{x}^2\,,
\end{equation}
where $a(t)$ is the scale factor.
The 00 component of Eq.~\eqref{eq:field} gives the modified Friedmann equation,
\begin{equation}
\label{firstFriedmann}
3 f_R H^{2}=\frac{1}{2}\left(f_R R-f\right)-3H\dot{f_R}+\kappa^{2}\rho_{M}\,,
\end{equation}
while the trace of the linear combination of Eq.~\eqref{eq:field} leads to the modified 
Friedmann acceleration equation,
\begin{equation}
\label{secondFriedmann}
2 f_R \dot{H} = -\ddot{f_R} + H\dot{f_R} - \kappa^{2}\left(\rho_{M}+P_{M}\right)\,,
\end{equation}
where the dot ``$\cdot$'' stands for the derivative with respect to the cosmic time $t$, $H
\equiv \dot{a} / a$ is the Hubble parameter, and $\rho_M = \rho_r + \rho_m$ ($P_M = P_r + P_m$) 
represent  the energy density (pressure) of relativistic ($r$) and non-relativistic ($m$) fluids.
Comparing with the origin Friedmann equations, we can get the  dark energy density and 
pressure as follows:
\begin{eqnarray}
&&  \rho_{DE} =\kappa^{-2}\left( \frac{1}{2}\left(f_R R-f\left(R\right)\right)-3H\dot{f_R}+3\left(1-f_R\right)H^2 \right)\,, \\
&& P_{DE} = \kappa^{-2}\left( -\frac{1}{2}\left(f_R R-f(R)\right)+\ddot{f_R}+2H\dot{f_R}-\left(1-f_R\right)\left(2\dot{H}+3H^2\right)\right) \,.
\end{eqnarray}
The equation of state of dark energy is defined by
\begin{eqnarray}
w_{DE}={ \rho_{DE} \over  P_{DE}}\,.
\end{eqnarray}
Following the same procedures in Refs.~\cite{Hu:2007nk, Bamba:2010ws}, we can simplify Eqs.~\eqref{firstFriedmann}
 and \eqref{secondFriedmann} 
to a second order differential equation,
\begin{eqnarray}
\label{diffeqyh}
y_{H}^{\prime \prime} + J_{1} y_{H}^{\prime} + J_{2}y_{H} + J_{3} = 0 \,,
\end{eqnarray}
with
\begin{eqnarray}
\label{yhyr}
y_{H}&\equiv& \frac{\rho_{DE}}{\rho_{m}^{(0)}}=\frac{H^{2}}{m^{2}}-a^{-3}- \chi a^{-4}\,, 
\nonumber\\
 J_{1}&=&4+\frac{1}{y_{H}+a^{-3}+\chi a^{-4}}\frac{1 - f_R}{6m^{2}f_{RR}} \,, ~~
J_{2}=\frac{1}{y_{H}+a^{-3}+\chi a^{-4}}\frac{2-f_R}{3 m^{2}f_{RR}} \,, 
\nonumber\\
 J_{3}&=&-3a^{-3}-\frac{\left( 1 - f_R \right)\left(a^{-3}+2\chi a^{-4}\right)+\left(R-f\right)/3 m^{2}}{y_{H}+a^{-3}+\chi a^{-4}}\frac{1}{6 m^{2}f_{RR}} \,,
\end{eqnarray}
where $m^{2}\equiv \kappa^2 \rho_{m}^{0}/3$, $\rho_i^{0} \equiv \rho_i (z=0)$, 
and $\chi \equiv \rho_{r}^{0} / \rho_{m}^{0}$, with $\rho_{m(r)}^{0}$ being the energy density
of the relativistic (non-relativistic) fluid at the present time.
Here, the prime ``$\prime$'' in Eq.~(\ref{diffeqyh}) denotes the derivative with respect to $\ln a$.
Using the differential equation in Eq.~(\ref{diffeqyh}), the cosmological evolutions can be calculated through the 
various existing programs in the literature.
Consequently, the deceleration parameter $q$ is found to be
\begin{eqnarray}
q &\equiv& -\left(1+{\dot{H}\over H^2}\right)
=\frac{1}{2}\frac{a^{-3}+2\chi a^{-4}+(1+3w_{\text{DE}})y_{H}}{a^{-3}+\chi a^{-4}+y_{H}}.
\end{eqnarray}

As the TM is one of the popular viable $f(R)$ gravity models, the conditions for  the viability must be satisfied.
For example, the TM has the following viable properties: (a) when $\lambda < cosh^{2}(R/R_{ch})$, $f_R=1-\lambda cosh^{-2}(R/R_{ch})>0$, leading to 
 a positive effective gravitational coupling;
 (b) when $\lambda >0$, $f_{RR} > 0$, resulting in a stable cosmological perturbation and a positivity of the  gravitational wave for the scalar;
(c) when $R\rightarrow \infty$, $f(R) \rightarrow R - 2 \Lambda$ with $\Lambda=\lambda R_{ch}/2$, showing 
an asymptotic behavior to the $\Lambda$CDM model in the large curvature region; 
and (d) when $\lambda> 0.905$, $0<m(R=R_d)<1$, indicating the existence of   a late-time stable de-Sitter solution,
where $m=Rf_{RR}/f_R$.

\section{Cosmological evolutions in Tsujikawa model}\label{sec:3}
\begin{figure}[]
\centering
\includegraphics[width=0.7 \linewidth]{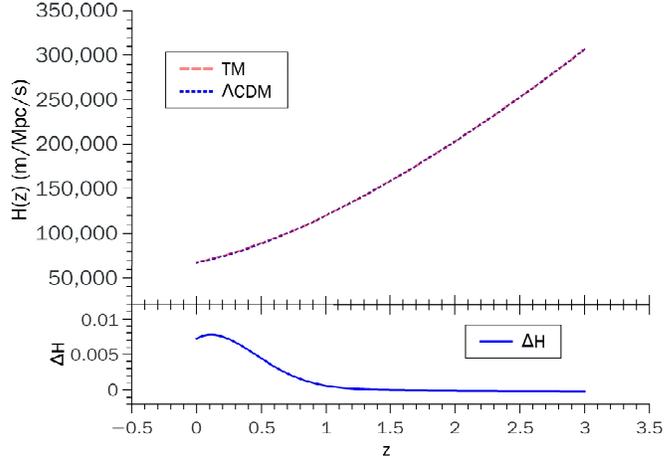}
\vskip 0 pt
\caption{Hubble parameter $H(z)$ as a function of the redshift $z$,
where the dashed (red) and dotted (blue) lines represent the TM and $\Lambda$CDM
for $\lambda^{-1}= 0.665$,
with $(H_{0,TM},H_{0,\Lambda CDM})=(67.62,67.71)$ $km/s \cdot Mpc$, 
and the initial conditions are given by $(\Omega_m, \Omega_r, \Omega_{DE} )_{TM}= (0.309, 7.88 \times 10^{-5}, 0.690) $ 
and $(\Omega_m, \Omega_r, \Omega_{DE} )_{\Lambda CDM}= (0.306, 7.88 \times 10^{-5}, 0.693) $,
respectively, while $\Delta H= (H_{TM}-H_{\Lambda CDM})/H_{\Lambda CDM}$. }
\label{fig:1} 
\end{figure}

\begin{figure}[]
\centering
\includegraphics[width=0.7 \linewidth]{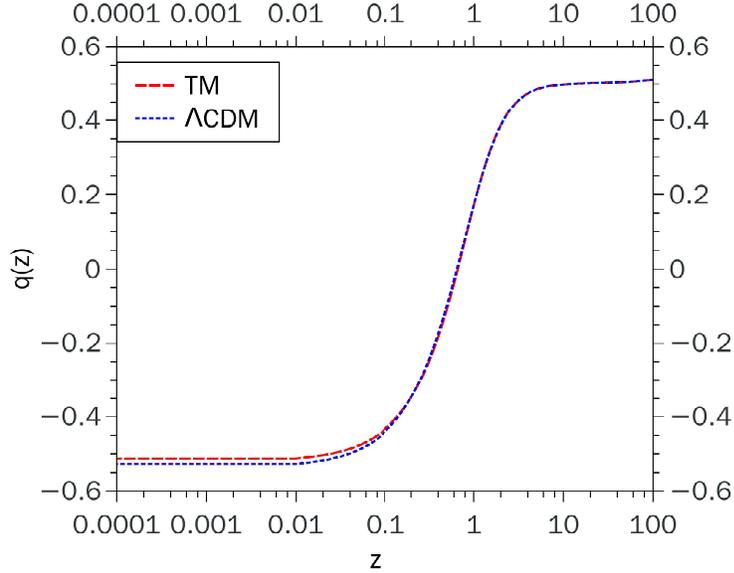}
\vskip 0 pt
\caption{Deceleration parameter $q(z)$ as a function of the redshift $z$,
where the legend is the same as  Fig.~\ref{fig:1}.}
\label{fig:2}
\end{figure}

To explore the expansion history and the linear perturbation of the universe in the TM, we use the MGCAMB program. 
In particular, we examine the cosmological parameters in 
the evolutions of the universe with the TM of  viable $f(R)$ gravity.
The initial conditions for the model are from the MGCosmoMC fitting, in which 
the input parameters have been chosen as the mean values.
In Figs.~\ref{fig:1} and \ref{fig:2}, 
we show  Hubble and deceleration parameters for the TM and $\Lambda$CDM, respectively.
We see that the difference between the two models in Fig.~\ref{fig:1}  is less than
$1\% $ in the whole expansion history of the universe. There are two reasons.
The first one is that the initial energy density ratios of matter and dark energy 
in the two models are close to each other.
The second one is that the TM is a $\Lambda$CDM-like theory, in which 
 it gives only a tiny contribution to the total energy density  before the dark energy dominated era.
 In Fig.~\ref{fig:2}, the behaves of the deceleration parameter in the two models are similar 
when $ z > 0.2 $.
The TM starts to have an accelerated expansion of the universe at $z = 0.688 $,
compared with $ z = 0.649 $ in the $\Lambda$CDM model.
In the present time, the difference is within $0.3\%.$
Clearly, it is hard to distinguish these two models by  either $H$ or $q$.

\begin{figure}[]
\centering
\includegraphics[width=0.65 \linewidth]{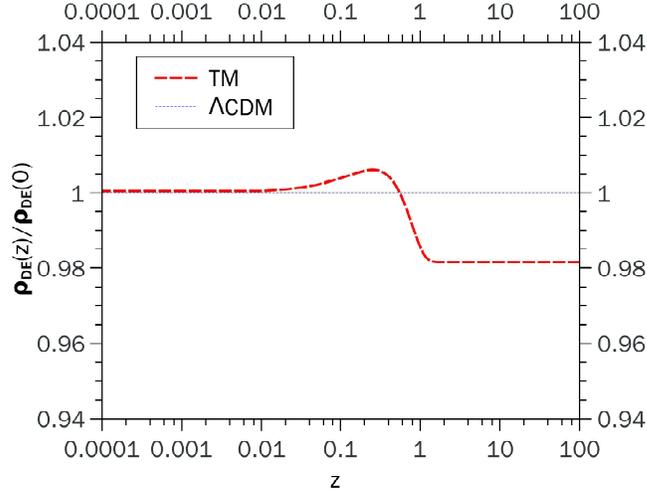}
\vskip 0 pt
\caption{Evolutions of the normalized effective dark energy density $\rho_{DE}(z)/\rho_{DE}(0)$ in the TM and
 $\Lambda$CDM. }
\label{fig:3}
\end{figure}

\begin{figure}[]
\centering
\includegraphics[width=0.65 \linewidth]{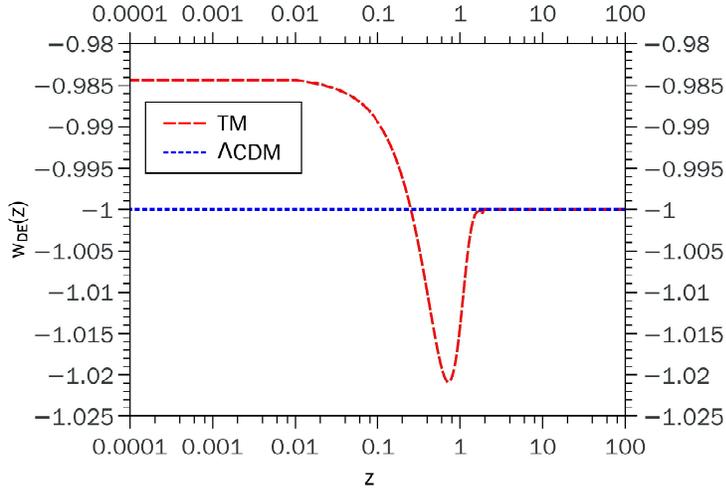}
\vskip 0 pt
\caption{Equation of state $w(z)$ for dark energy as a function of $z$  in the TM and
 $\Lambda$CDM. }
\label{fig:4}
\end{figure}

As one of the characteristics in the viable $f(R)$ models, the behavior of the TM
 approaches the cosmological constant when $z$ is large. In Fig.~\ref{fig:3}, the effective
 energy density is almost constant in the early time, which is smaller than the present dark energy density. 
When $z<1.0$, it starts to rise and fall slightly. The equation of state evolution is shown in  Fig.~\ref{fig:4}. 
For the TM, it indeed oscillates and crosses the phantom divide line as mentioned in Ref.~\cite{Bamba:2010iy}.

\begin{figure}[]
\centering 
\subfigure{
\label{fig.sub.3}
\includegraphics[width=0.45\textwidth]{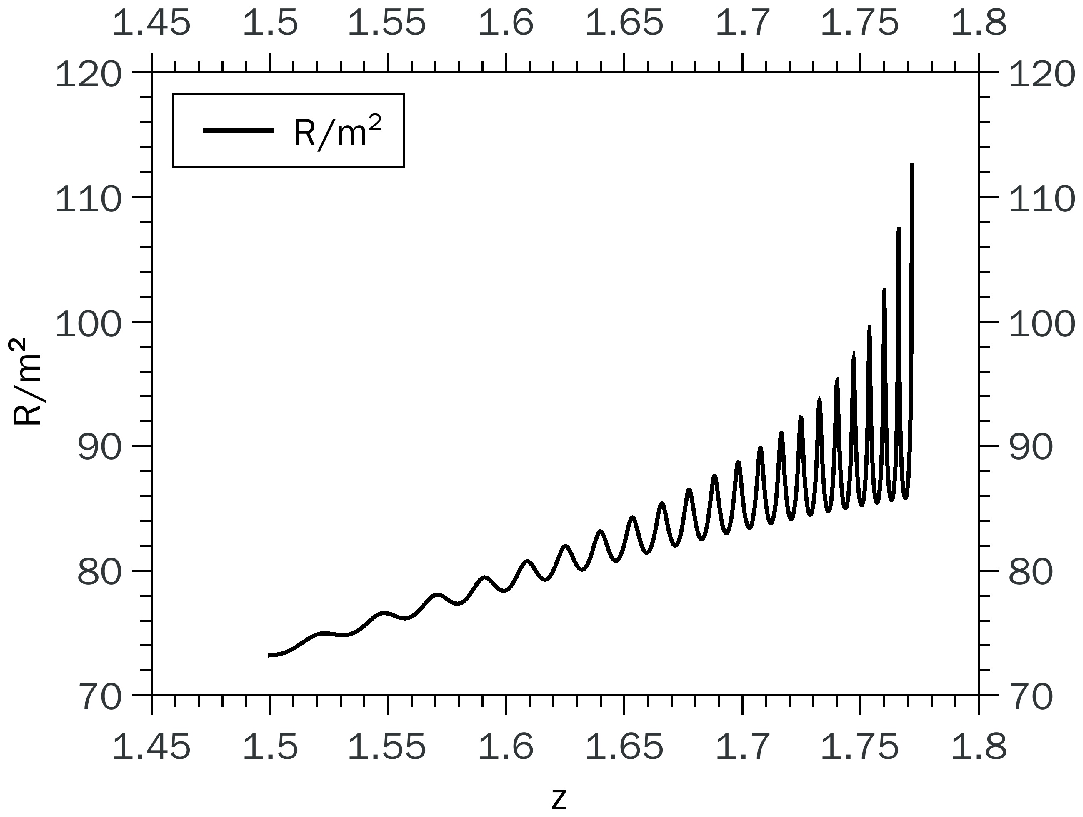}}
\subfigure{
\label{fig.sub.4}
\includegraphics[width=0.50\textwidth]{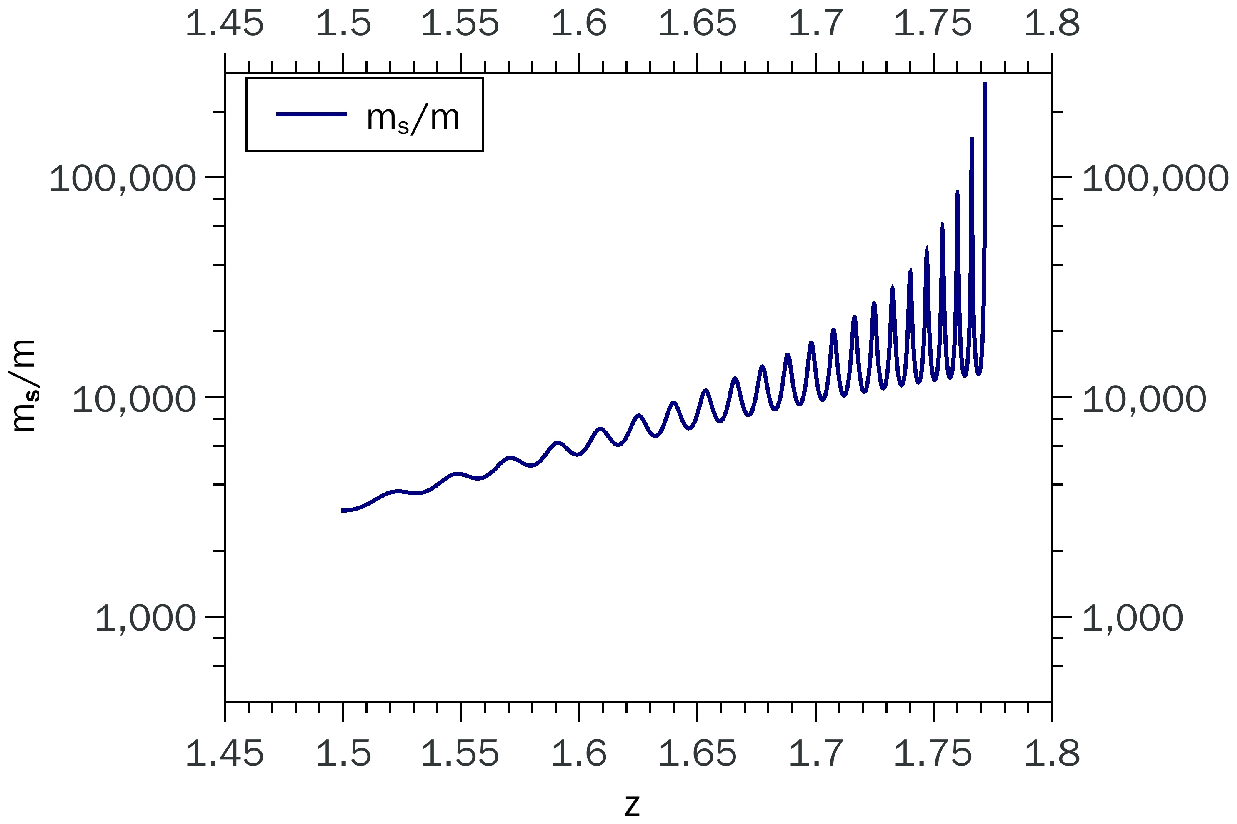}}
\caption{Normalized Ricci Scalar $R/m^{2}$ and scalaron mass $m_s/m$ as the functions of $z$ in the TM.}
\label{fig.5}
\end{figure}

In Fig.~\ref{fig.5}, we show the cosmological evolutions of the normalized Ricci scalar  $R/m^2$ and scalaron mass $m_s/m$
as the functions of $z$ in the TM with $m\equiv \kappa (\rho_{m}^{0}/3)^{1/2}$.
In the cosmological background evolution, the singularity problem is not avoidable 
because of the generic  property of the viable $f(R)$ models. As $z$ gets larger, the 
scalaron mass becomes much heavier so that the Ricci scalar strongly 
oscillates, which causes the program to fail easily. By considering the 
asymptotic behavior of the $\Lambda$CDM model in the viable conditions and solving
 the differential equation in the z decreasing direction, the numerical error can be 
handled in some code technique. The other way is to add the $R^2$  term in the $f(R)$ 
models as mentioned in Ref.~\cite{Lee:2012dk}. This term can also help the 
models to have a steady performance in the $R$ evolution. 

From Ref.~\cite{Yang:2011cp}, we know that the scalaron mass also affects
 the propagation of the scalar mode of the gravitational waves. It will make the
mode to decay so fast below the cutoff frequency in the viable $f(R)$ models. From 
the numerical result, we can obtain the cutoff frequency in the background 
around $10^{-17}$~Hz now in the TM.
 If the wave propagates in the more dense region, such as the inner galaxy with the density around 
$10^{-24} g/cm^{3} $, the frequency will rise to infinity. This may be tested in the future stochastic gravitational wave detection.

\begin{figure}[]
\graphicspath{ {./Figures/} }
\centering
\includegraphics[width=0.7 \linewidth]{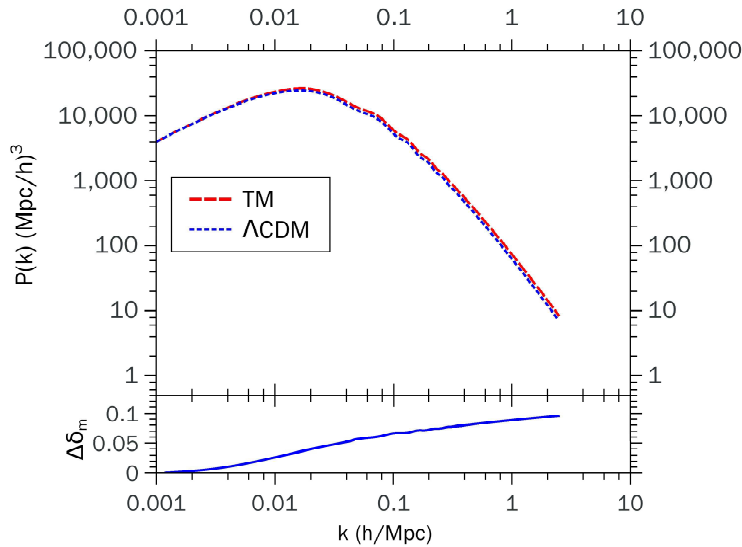}
\vskip 0 pt
\caption{Spectra of the matter power perturbation in the TM and $\Lambda$CDM,
where $\delta_m$ is the matter density of the perturbation and $\Delta\delta_{m} = (\delta_m^{TM}-\delta_{m}^{\Lambda CDM})/\delta_{m}^{\Lambda CDM}$, 
.}
\label{fig.6}
\end{figure}

\begin{figure}[]
\centering
\includegraphics[width=0.7 \linewidth]{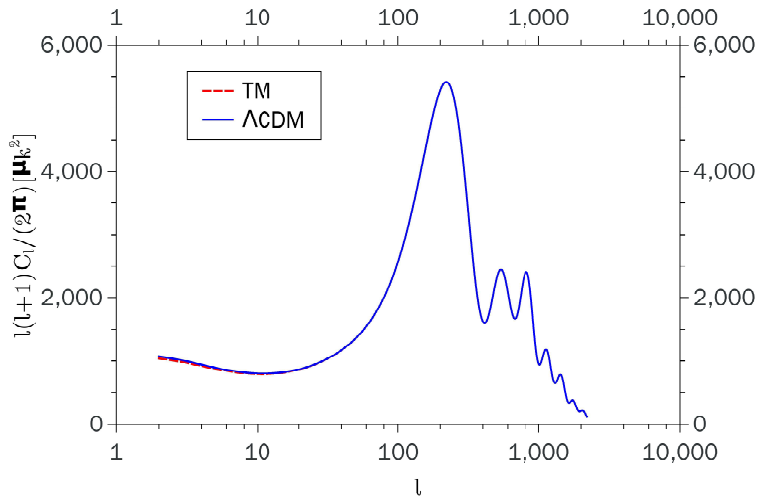}
\vskip 0 pt
\caption{Spectra for the cosmic microwave background in the TM and  $\Lambda$CDM. }
\label{fig.7}
\end{figure}

The TM can be seen as the same branch of the exponential model of viable $f(R)$ gravity. Compared with
the TM,  the background evolution in the exponential model illustrates a more 
sharp variation. The related work has been done in Ref.~\cite{CQG:2019}.
We can compare these two models in the linear perturbation theory.
In the original CAMB program, we choose the synchronous gauge to do the simulation.
But in the open source of MGCAMB, the Newtonian gauge is used to do the calculation,
in which the metric  is given by~\cite{Tsujikawa:2007gd, Ma:1995ey}
\begin{eqnarray}
\label{eq:metric}
ds^{2}=-\left(1+2\Psi\right)dt^{2}+a^{2}\left(t\right)\left(1-2\Phi\right)d\vec{x}^{2} \,.
\end{eqnarray}
Under the subhorizon limit, one has that~\cite{CQG:2019}
\begin{eqnarray}
\label{eq:reducedperturb}
 \frac{k^2}{a^2}\Psi = -4\pi G \mu\left(k,a\right)\rho_M \Delta_M  
\end{eqnarray}
with
\begin{align}
\label{eq:parmugamma}
\mu\left(k,a\right) 
= \frac{1}{f_R}\frac{1+4\frac{k^2}{a^2}
  \frac{f_{RR}}{f_R}}{1+3\frac{k^2}{a^2}
         \frac{f_{RR}}{f_R}} 
\quad\text{and}\quad
\frac{\Phi}{\Psi} = \gamma\left(k,a\right) 
= \frac{1+2\frac{k^2}{a^2}
  \frac{f_{RR}}{f_R}}{1+4\frac{k^2}{a^2}\frac{f_{RR}}{f_R}} \,,
\end{align}
where $k$ is the comoving wavenumber and $\Delta_M \equiv \delta_{M}+3 H\left(1+\omega_M\right)v_M/k$ 
is the gauge-invariant matter density perturbation with $w_M = P_M / \rho_M$  the equation of state  and $v_M$  the velocity for matter.
 The growth equation for the matter density perturbation at the matter dominated epoch with $v_m = 0$ is given by
\begin{eqnarray}
\label{eq:deltam}
\ddot{\delta}_m + 2H\dot{\delta}_m - 4 \pi G \mu(k,a) \rho_m \delta_m = 0 \,.
\end{eqnarray}
 As shown in Sec.~II, the TM satisfies the viable conditions of $0 < f_R < 1$ and $f_{RR} > 0$,
which imply two scenarios. Firstly, if $k$ increases, it will cause a larger value of $\mu(k,a)$. Secondly,  the matter density fluctuations are enhanced due to a larger value of $\mu(k,a)$, which is regarded as the scale independent and dependent factors of $f_R^{-1}$ and $\left( 1+4k^2f_{RR}/(a^2f_R)\right) / \left(1+3k^2f_{RR}/(a^2f_R) \right)$ 
for $k^2 \ll f_{RR}/(a^2f_R)$ and  $k^2 \gg f_{RR}/(a^2f_R)$, respectively.
Besides, if we consider the wavenumber outside the Hubble radius, i.e. $k \rightarrow 0$, the scalar perturbation will obey~\cite{Song:2006ej}
\begin{align}
\label{phi_evolution}
\Phi''+\left(1-\frac{H''}{H'}+\frac{B'}{1-B}+B \frac{H'}{H}\right)\Phi'&+\left(\frac{H'}{H}-\frac{H''}{H'}+\frac{B'}{1-B}\right)\Phi=0\, ,    (k=0).\, \\
\label{poisson}
2\Phi_{eff}+\frac{B}{2}\frac{E'}{E}&\frac{E'}{4E'+E''}S=\frac{-1}{f_R}\frac{\kappa^2a^2\rho_M}{k^2}\Delta_M\
\end{align}
where
\begin{align}
\label{b_eqn}
 B     = \frac{f_{RR}}{f_R} R'\frac{H}{H'}\,, \quad E= \frac{H^2}{H_0^2},\, \quad 
 \Phi_{eff}=\frac{1}{2}(\Phi+\Psi)\,,  \quad S= -2\Phi+\Psi\,.
\end{align}
 In the TM and $\Lambda$CDM, 
 in terms of  Eq.~(\ref{phi_evolution}), $\Phi$ grows as $a$ increases. However, the change rate in the TM is higher than that in the $\Lambda$CDM model. 
 With the relation $\Psi=(B\Phi'+\Phi)/(1-B)$, one obtains 
   a smaller value for the effective gravitational potential~\cite{Song:2006ej}. Due to the negative sign of  the second term in the LHS of 
   the modified Poisson equation in Eq.~(\ref{poisson}), the smaller negative $\phi_{eff}$ would finally enhance the matter 
density perturbation in the TM.
In Fig.~\ref{fig.6}, we concentrate on the sub-horizon regime, which is more relevant to the observation. 
Here, $k$  should be larger than 0.001 to  satisfy the sub-horizon limit of $k/aH \gg 1$.
The result in the TM has a more enhancement than that in the $\Lambda$CDM model
 for the higher value of $k$. 
 It also relaxes the limit of the neutrino mass sum a little  because there is more freedom for a larger value of $\Sigma m_\nu$ resulting 
 from  the suppressed effect. On the other hand, the viable modified gravity models also
 affect the CMB spectrum through the late-time integrated Sachs-Wolfe effect as shown in Fig.~\ref{fig.7}.
As the gravitational potential in the TM evolves in different ways comparing with $\Lambda$CDM, 
the figure in the TM declines slightly but still approaches to the $\Lambda$CDM result when $\ell<10$. 
However, for the observational data in this regime, the errors are still large.  
Strict constraints on the TM of modified gravity can be given from the CMB when more precise
measurements of the low-$\ell$ regime are available. 

\section{Constraints from Cosmological Observations}\label{sec:4}

\begin{figure}[!htb]
\graphicspath{ {./Figures/} }
\centering
\includegraphics[width=0.7 \linewidth]{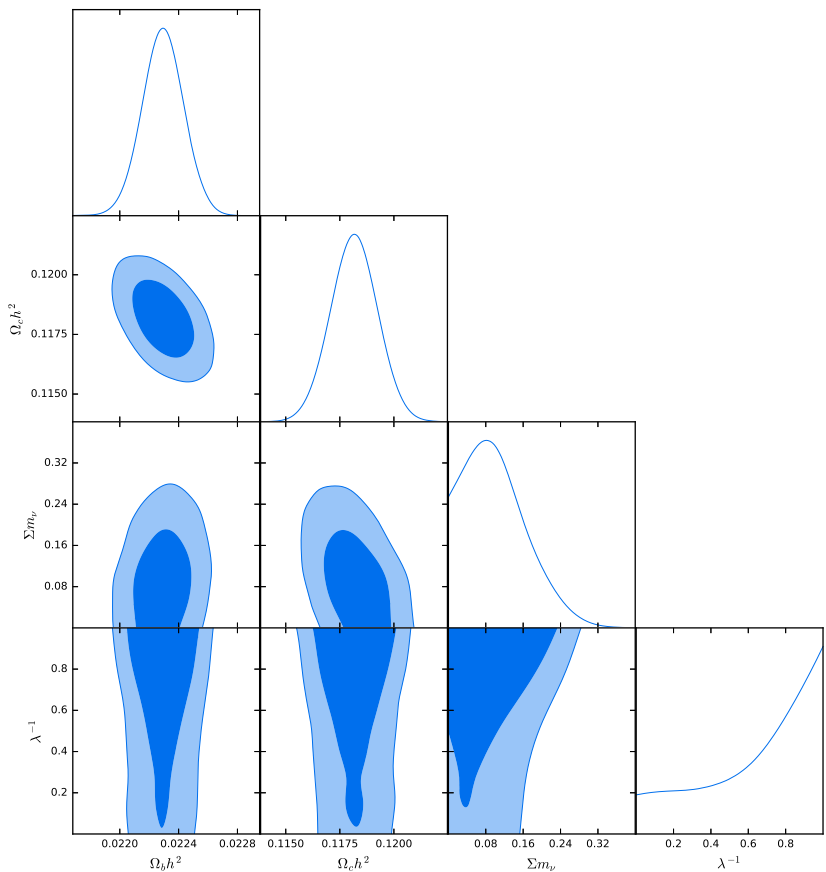}
\vskip 0 pt
\caption{Two-dimensional contour plots of $\Omega_b$, $\Omega_c$, $\lambda^{-1}$ and $\Sigma m_{\nu}$ in the TM. }
\label{fig:8}
\end{figure}

We have used the best fitted values from the MGCosmoMC to evaluate 
the background evolutions in the previous section. Now we 
would study the constraints from the cosmological observations.
With the MGCosmoMC, the input parameters are given in 
Table.~\ref{table:1} and the fitting results are shown in Table.~\ref{table:2}. 
In the following, we compare the results in the TM and $\Lambda$CDM.
 First,
the best fitted parameters in the TM are close to those in the $\Lambda$CDM model.
For example, $\Omega_bh^2_{bestfit}$ and $\Omega_ch^2_{bestfit}$ are almost the same ($<0.1\%$)
in the two models. 
However, there are also some parameters in the TM, which can deviate from the $\Lambda$CDM. 
In particular, we find that the limit of the neutrino mass sum $\Sigma m_\nu$ in the TM
is about $20\%$ larger than that in the  $\Lambda$CDM one.
This result extends the discuss  in the 
matter power spectrum for  $\Sigma m_\nu$. 
Finally, for the contour plots in  Fig.~\ref{fig:8}, we can see 
that most of the results in the TM are similar to those in the exponential model 
mentioned in  Ref.~\cite{CQG:2019}, except the model parameter. Note that the model parameter is more
 sensitive in the exponential model because its 1$\sigma$ range is obviously smaller than that in the TM. 
\begin{table}
\caption{List of priors} \label{table:1}
\begin{tabular}{|c|c|} \hline
Parameters & Priors
\\ \hline\hline
Model parameter & $10^{-4}< \lambda^{-1} <1$
\\ \hline 
Baryon density & $5 \times 10^{-3}<\Omega_bh^2<0.1$
\\ \hline
CDM density & $10^{-3}<\Omega_ch^2<0.99$
\\ \hline
Neutrino mass & $0<\Sigma m_{\nu} < 1 $ eV
\\ \hline
Spectral index & $ 0.9 < n_s < 1.2$
\\ \hline
Scalar power spectrum amplitude & $ 2 <\mathrm{ln}(10^{10} A_s) < 4$
\\ \hline
Reionization optical depth & $ 0.01 <\tau <0.8$
\\ \hline
$100 \ \theta_{\text{MC}}$  & $ 0.5 <100 \ \theta_{\text{MC}} <10$
\\ \hline
Hubble parameter (km/s $\cdot$ Mpc)  & $ 20 < H_0 < 100$
\\ \hline

\end{tabular}
\end{table}

\begin{table}[h!]
\caption{Fitting results in TM and $\Lambda$CDM }\label{table:2}
\begin{tabular}{|l|c|c|} 
 \hline
Parameters & TM & $\Lambda$CDM \\
\hline \hline
$\lambda^{-1}$ & $0.6646_{-0.54362}^{+0.33544} $ & -
\\ \hline
$\Omega_bh^2$ & $0.02229_{-0.00028}^{+0.00028}$   & $0.02229_{-0.00027}^{+0.00027}$
\\ \hline
$\Omega_ch^2$ & $0.11816_{-0.00212}^{+0.00212}$ & $0.11816_{-0.00215}^{+0.00211}$
\\ \hline
$\Sigma m_{\nu}$  &$ 0.10392_{-0.10392}^{+0.12283}$ & $0.08434_{-0.08434}^{+0.11816}$
\\ \hline
$ n_s$ & $0.96867_{-0.00763}^{+0.00756}$ & $0.96899_{-0.00770}^{+0.00772}$
\\ \hline
$\mathrm{ln}(10^{10} A_s)$ & $3.06229_{-0.05305}^{+0.05364}$  & $3.07018_{-0.05092}^{+0.05484}$
\\ \hline
$\tau$ & $0.06630_{-0.02789}^{+0.02892}$ & $0.07012_{-0.02690}^{+0.02944}$
\\ \hline
$100 \ \theta_{MC}$  & $1.04090_{-0.00059}^{+0.00059}$ & $ 1.04090_{-0.00059}^{+0.00059}$
\\ \hline
$ H_0$  $(km/s \cdot Mpc)$  & $67.62788_{-1.21858}^{+1.13616}$   & $67.71428_{-1.25371}^{+1.11201}$
\\ \hline
$ {\rm{Age}}/{\rm{Gyr}}$  & $13.81232_{-0.06538}^{+0.07030}$ &  $13.81044_{-0.06323}^{+0.07023}$
\\ \hline
$\sigma_8$  & $0.85866_{-0.05717}^{+0.04351}$ & $0.81101_{-0.02710}^{+0.02411}$
\\ \hline
$ \chi^2_{best-fit}$  & 13458.82 &  13459.12\\
 \hline
\end{tabular}
\end{table}

\section{Conclusions}\label{sec:5}

We have studied the cosmological evolutions of the universe in the TM of viable f(R) gravity, 
which  have also been compared with those in the $\Lambda$CDM model. 
We have found that the results in the TM are not much different from the corresponding ones in the $\Lambda$CDM.
We have demonstrated that the transition point from the deceleration to acceleration in our universe is $z= 0.688$ in the TM,
 which is higher than $z = 0.649$ in the  $\Lambda$CDM model. As a result, the dark energy dominance is slightly pushed up in the TM. 

For the large scale structure, the amplitude of the matter power spectrum in the TM is  strenghen in $ k>0.0013$. 
At $k =0.2$ of the linear perturbation limit, the amplitude is about $3.9\%$ larger than that in the $\Lambda$CDM.
On the other hand, the TM of viable $f(R)$ gravity affects the gravitational potential evolution through the modified Possion equation.
In the CMB spectrum, when $\ell<10$, it is sensitive to the change of 
 gravitational potential in the
universe history. There is only slightly difference between the TM and $\Lambda$CDM
in this region, whereby both of them fit very well in other regions. 
As the current observational data are not accurate enough, it is still possible 
 to test GR and modified gravity models when more future measurements are available.
From the contour plots for the parameter fittings, we have displayed that the TM also gives a relaxed constraint on the
neutrino mass sum as the other viable $f(R)$ gravity models. In addition,  the model parameter in the TM is more sensitive than 
that in the exponential model.
Our numerical results have demonstrated some different features among the viable $f(R)$ models.
If we  fully clarify the characters of these models and estimate their cosmological evolutions,
they can potentially  hint about what is next to do in the dark energy research in the future.

\section*{ACKNOWLEDGMENTS}
We thank Dr. Ling-Wei Luo for some useful discussion.
This work was supported in part by National Center for Theoretical Sciences and 
MoST (MoST-104-2112-M-007-003-MY3 and MoST-107-2119-M-007-013-MY3).

  \end{document}